# Bulk moduli of $PbS_xSe_{1-x}$, $PbS_xTe_{1-x}$ and $PbSe_xTe_{1-x}$ from a thermodynamical model compared to generalized gradient approximation approach.

E. S. Skordas*

Department of Solid State Physics, Faculty of Physics, University of Athens,
Panepistimiopolis, 157 84 Zografos, Greece

**Abstract**

Very recently, first-principle technique of full-potential linearized augmented plane-wave method, by using for exchange-correlation potential the generalized gradient approximation (GGA), was employed for the study of the lead chalcogenide semiconductors' alloys $PbS_xSe_{1-x}$, $PbS_xTe_{1-x}$ and $PbSe_xTe_{1-x}$. These density functional calculations led to the determination of structural, electronic and optical properties, including the values of lattice constants and bulk moduli as a function of composition. Here, we investigate the latter properties, but by employing a thermodynamical model which has been suggested for the formation and migration of defects in solids including several recent applications in semiconductors. The following crucial difference emerges when comparing the present results with those deduced by density functional calculations: Among the alloys studied, GGA calculations identify that $PbS_xTe_{1-x}$ exhibits the most evident non-linear variation of the bulk modulus versus the composition, while according to the thermodynamical model such an evident non-linear behavior –and maybe somewhat stronger- is also expected for $PbSe_xTe_{1-x}$. A tentative origin of this diversity is discussed.

# 1. Introduction

A lot of experimental work has been done for the study of the structural [1, 2], electronic [3, 4] and optical properties [5, 6] of Lead chalcogenide semiconductors that exhibit rock salt structure. The fact that they have small fundamental energy bandgap [7, 8] make them very useful in optoelectronic equipments including lasers and detectors [9-12]. For example, they have been extensively used as thermoelectric materials, infrared detectors, in window coating and in panels used for solar energy utilization. Furthermore, they are very promising for the photoinduced nonlinear optics [13, 14] and in all the cases their bulk properties are crucial due to substantial contributions of anharmonic phonons.

These important materials have been also studied by various theoretical techniques [15-18]. Very recently the structural, electronic and optical properties of $PbS_xSe_{1-x}$, $PbS_xTe_{1-x}$ and $PbSe_xTe_{1-x}$ ternary alloys have been studied by Naeemullah et al [19] by using first-principle technique of full-potential linearized augmented plane-wave (FP-LAPW) method. In these density functional calculations, Naeemullah et al used the FP-LAPW method with Wu-Cohen generalized gradient approximation (GGA) [20] in order to solve Kohn-Sham equation [21]. The wave function, potential and change density were expanded into two different bases: The wave function was expanded in spherical harmonics in the atomic spheres while outside the spheres (i.e., interstitial regions) it was expanded in plane wave basis. The potential was also expanded in the same manner.

Naeemullah et al [19] deduced results for the structural, electronic and optical properties of the alloys $PbS_xSe_{1-x}$, $PbS_xTe_{1-x}$ and $PbSe_xTe_{1-x}$ for $x$=0.25, 0.5, 0.75 and 1. Here, we shall restrict ourselves to the structural properties. In particular,

Naeemullah et al calculated values of the equilibrium lattice constant ($\alpha$), bulk modulus (*B*), the pressure derivative of the bulk modulus ($B' \equiv \left.\frac{dB}{dP}\right|_T$) and the minimum energy *E* for binary compounds. By plotting their calculated lattice parameters versus the composition *x,* Naeemullah et al found the following: For $PbS_xSe_{1-x}$ and $PbSe_xTe_{1-x}$ alloys they exhibit tendency to Vegard's law [22] (which predicts almost linear variation versus *x*) with a marginal upward bowing parameters equal to -0.032 and -0.079 Å, respectively. On the other hand, for the $PbS_xTe_{1-x}$ alloy, a large deviation from Vegard's law with upward bowing parameter equals to -0.2125Å was observed [19]. As for the variation of the bulk modulus versus composition, their results showed a significant deviation from the linear concentration dependence with downward bowings equal to 6.673 and 4.08 GPa for $PbS_xTe_{1-x}$ and $PbSe_xTe_{1-x}$, respectively. On the other hand, only a small deviation from Vegard's law was found for $PbS_xSe_{1-x}$ alloy with a downward bowing parameter equal to 0.983 GPa. In other words, as far as the values of the bulk modulus of the aforementioned three alloys is concerned, Naeemullah et al's density functional calculations showed that the system $PbS_xTe_{1-x}$ exhibited the strongest deviation (i.e., 6.673 GPa) from linearity.

More recently, however, a pioneering experimental study of the compressibility of Ir-Os alloys appeared by Yusenko et al. [23]. After preparing several fcc- and hcp- structured Ir-Os alloys Yusenko et al studied these alloys up to P = 30GPa at room temperature by means of synchrotron-based X-ray powder diffraction in diamond anvil cells and found that their bulk moduli increase with increasing osmium content showing a deviation from linearity. This concentration dependence of bulk moduli was shown [23] to be satisfactorily described by means of a thermodynamical model [24] for the formation and migration of the defects in solids [25] which has been successfully applied to a variety of solids including metals [26], alloys [27], fluorides [28], mixed alkali halides [29], semiconductors [30, 31, 32, 33], actinide oxides particularly useful for nuclear fuel applications [34, 35] as well as complex materials under uniaxial stress that emit electric signals before fracture [36], which explains the signals detected before major earthquakes [37, 38]. It is therefore challenging to apply here this model to the case of the three alloys $PbS_xSe_{1-x}$, $PbS_xTe_{1-x}$ and $PbSe_xTe_{1-x}$ and compare the results with those deduced from the density functional calculations by Naeemullah et al [19].

**2. The model that interrelates the compressibility of an alloy with the compressibilities of its end constituents.**

We consider for simplicity an alloy resulted from two pure components (1) and (2). Let $\upsilon_1$ be the volume per "molecule" of the pure component (1) and $\upsilon_2$ the volume per "molecule" of the pure component (2), thus the molar volumes for the two end members are $V_1 = N\upsilon_1$ and $V_2 = N\upsilon_2$, where $N$ stands for Avogardos's number. By considering the volume $V$ of the alloy in which $n$ "molecules" have replaced $n$ molecules of the pure body, the molar fraction $x$ is connected to $n/N$ by $n/N = x/(1-x)$. Then, upon assuming that each replacement of a "molecule" constitutes a defect that affects the volume of the system by the same amount independent of its composition, we may write to a first approximation [24]

$$V = (1-x)V_1 + xV_2 \qquad (1)$$

By differentiating Eq.(1) in respect to pressure $P$ and considering the compressibilities $\kappa_1 = -\frac{1}{V_1}\frac{dV_1}{dP}\Big|_T$ and $\kappa_2 = -\frac{1}{V_2}\frac{dV_2}{dP}\Big|_T$ for the two pure constituents (1) and (2) respectively, we get for the compressibility $\kappa\left(=-\frac{1}{V}\frac{dV}{dP}\Big|_T\right)$ of the alloy:

$$\kappa V = (1-x)\kappa_1 V_1 + x\kappa_2 V_2 \tag{2}$$

By inserting Eq.(1) into Eq.(2) we finally get [24]

$$B = B_1 \frac{1 + x\left(\frac{V_2}{V_1} - 1\right)}{1 + x\left(\frac{B_1 V_2}{B_2 V_1} - 1\right)} \tag{3}$$

where $B_1\left(=\frac{1}{\kappa_1}\right)$ and $B_2\left(=\frac{1}{\kappa_2}\right)$ and the bulk moduli for the pure constituents (1) and (2) and $B\left(=\frac{1}{\kappa}\right)$ for the bulk modulus for the alloy. Note that Eq.(3) stems from Eq.(1), thus the applicability limits of Eq.(3) depend on whether the assumption that led to Eq.(1) is valid. The following comments are now in order:

First, Eq.(3) enables the direct evaluation of $B$ at any desired composition in terms of the elastic data of the two end members.

Second, Eq.(3) in general points to a non linear variation of the bulk modulus of the alloy versus the composition $x$.

Third, in the latter case, i.e., when the measurements show a non linear variation of $B$ versus $x$, and in addition the values $B_1$ and $B_2$ are not accurately known (e.g., due to experimental difficulties), which was the recent case of Ir-Os alloys studied by Yusenko et al [23] mentioned above, Eq.(3) can be used as follows: The experimental bulk moduli for alloys can be fitted using Eq.(3), and the best fit may lead to reliable estimates of $B_1$ and $B_2$. This was followed by Yusenko et al [23] who found the values 354(2) GPa and 442(4) GPa for pure Ir and Os, respectively, which were in good agreement with experimental values obtained independently.

## 3. Application of the thermodynamical model to $PbS_xSe_{1-x}$, $PbS_xTe_{1-x}$ and $PbSe_xTe_{1-x}$ alloys.

Our recent results are given in Tables 1, 2 and 3 for the alloys $PbS_xSe_{1-x}$, $PbS_xTe_{1-x}$ and $PbSe_xTe_{1-x}$ respectively ($x$ =0, 0.25, 0.5, 0.75 and 1). Concerning these results we point out the following:

Equation (1) can be alternatively written as

$$\alpha^3 = (1-x)\alpha_1^3 + x\alpha_2^3 \qquad (4)$$

Where $\alpha$ is the lattice parameter for the alloy and $\alpha_1$, $\alpha_2$ the lattice parameters for the pure constituents (1) and (2). On the basis of Eq.(4) we can deduce of course an estimation of $\alpha$ for each composition by using the experimental values of $\alpha_1$ and $\alpha_2$ which have been reported by Denton et al [39], Dalven et al [7], Cohen and Chelikowsky [3] and Delin et al [15]. Although these experimental values for each compound differ only slightly among the various authors, we preferred here to use in our calculations in Eq.(4) the average of their published values which are: 5.936, 6.1237 and 6.461 Å for PbS, PbSe and PbTe, respectively. (Note, however, that our results are not practically affected if we use the experimental values published by the same authors.) The same values were also used for our calculations concerning $V_1$ and $V_2$ in Eq.(3). Finally, concerning the bulk moduli $B_1$ and $B_2$ in Eq.(3) we used the experimental values published by Denton et al [33] which are: 52.9, 54.1 and 39.8 GPa for PbS, PbSe and PbTe, respectively.

In Tables 1, 2 and 3 we give for each composition in the alloys $PbS_xSe_{1-x}$, $PbS_xTe_{1-x}$ and $PbSe_xTe_{1-x}$ the values of $\alpha$ and $B$ obtained by means of Eqs(4) and (3), respectively. The corresponding $B$ values are also plotted versus the composition in Fig. 1. In these Tables, for the sake of comparison, we also insert for each composition the corresponding values of $\alpha$ and $B$ deduced by Naeemullah et al [19]. This comparison reveals the following striking difference: It is clear from an inspection of Fig. 1 that, among the three types of alloys investigated here, $PbSe_xTe_{1-x}$ and $PbS_xTe_{1-x}$ exhibit evident non linear variation of $B$ versus $x$, which seems to be somewhat stronger in $PbSe_xTe_{1-x}$. (In other words, we find that for $PbSe_xTe_{1-x}$ alloys we do not observe tendency to Verard's law, i.e., $B$ does not exhibit linear variation versus $x$) On the other hand, in the calculated values by Naeemullah et al, as mentioned in the Introduction, the most intense nonlinear behaviour was found in $PbS_xTe_{1-x}$. Our findings here could be understood in the frame of the following context: In $PbSe_xTe_{1-x}$ the bulk moduli of its pure constituents PbSe and PbTe exhibit the largest difference since their experimental values, as mentioned, are 54.1 and 39.8 GPa respectively, while in $PbS_xTe_{1-x}$ the difference is smaller since the corresponding bulk moduli of its pure constituents are 52.9 and 39.8 GPa respectively. Hence, in the former case ($PbSe_xTe_{1-x}$), we intuitively expect a larger extent of the anharmonicity (thus, a larger deviation from Vegard's law) compared to the latter case.

## 4. Conclusion

Here, by employing a thermodynamical model which has been recently found [30-33] to give successful results in several applications of defects in semiconductors we calculated the dependence of the bulk modulus versus the composition in the alloys $PbS_xSe_{1-x}$, $PbS_xTe_{1-x}$ and $PbSe_xTe_{1-x}$. While the GGA calculations predict that the variation of the bulk modulus versus the composition exhibited the most evident deviation from linearity for the alloy PbS$_x$Se$_{1-x}$, the thermodynamical model reveals that such a behavior is also evident and may be somewhat stronger for PbSe$_x$Te$_{1-x}$. This, is a challenge for future experimental investigation.

**References**

[1] G. Nimtz, B. Schlicht, B. Dornhaus, Narrow Gap Semi-Conductors: Springer Tracts in Modern Physics, Springer, New York, 1983.

[2] K.M. Wong, W.K. Chim, J.Q. Huang, L. Zhu, J. Appl. Phys. 103 (2008) 054505.

[3] M.L. Cohen, J.R. Chelikowsky, Electronic Structure and Optical Properties of Semiconductors, Springer Series in Solids States Sciences, vol. 75, second ed,Springer, Verlag, Berlin, 1989.

[4] K.M. Wong, Jpn. J. Appl. Phys. 48 (2009) 085002.

[5] M. Cardona, D.L. Greenaway, Phys. Rev. 133 (1964) 1685.

[6] D. Korn, R. Braunstein, Phys. Rev. B 5 (1972) 4837.

[7] R. Dalven, H. Ehrenreich, F. Seitz, D. Turnbull, Solid State Physics, vol. 28, Academic, New York, 1973. pp. 179.

[8] R.A. Cowley, Philos. Mag. 11 (1965) 673.

[9] G. Springholz, V. Holy, M. Pinczolits, G. Bauer, Science 282 (1998) 734.

[10] E. Khoklov (Ed.), Lead Chalcogenides: Physics and Applications, Taylor and Francis, New York, 2003.

[11] M. Tacke, Infrared Phys. Technol. 36 (1995) 447.

[12] Z. Feit, M. Mc Donald, R.J. Woods, V. Archambauld, P. Mak, Appl. Phys. Lett. 66 (1996) 738.

[13] I.V.Kityk, M. Demianiuk, A. Majchrowski, J. Ebothe, P. Siemion, J.Phys.: Condens.Mater V. 16, (2004) 3533-3544.

[14] K. Nouneh, I. V. Kityk, R. Viennois, S. Benet, S. Charar, S. Paschen, K. Ozga, Phys. Rev. B 73 (2006) 035329.

[15] A. Delin, P. Ravindran, O. Eriksson, J.M. Wills, Int. J. Quantum Chem. 69 (1998) 349.

[16] M. Lach-hab, A.D. Papaconstantopoulos, J.M. Mehl, J. Phys. Chem. Solids 63 (2002) 833.

[17] E.A. Albanesi, C.M.I. Okoye, C.O. Rodriguez, E.L. Peltzer y Blanca, A.D. Petukhov, Phys. Rev. B 61 (2000) 16589.

[18] E.A. Albanesi, E.L. Peltzer y Blanca, A.G. Petukhov, Comput. Mater. Sci. 32 (2005) 85.

[19] Naeemullah, G. Murtaza, R. Khenata, N. Hassan, S. Naeem, M.N. Khalid, S. Bin Omran, Computational Materials Science 83 (2014) 496–503.

[20] Z. Wu, R.E. Cohen, Phys. Rev. B 73 (2006) 235116.

[21] W. Kohn, L.J. Sham, Phys. Rev. 140 (1965) A1133.

[22] L. Vegard, Z. Phys. 5 (1921) 17.

[23] K. V. Yusenko, E. Bykova, M. Bykov, S. A. Gromilov, A. V. Kurnosov, C. Prescher, V. B. Prakapenka, M. Hanfland, S. van Smaalen, S. Margadonna, et al., J. Alloys Compd. 622 (2015) 155.

[24] P. Varotsos, Phys. Status Sol. B 99 (1980) K93;

P. Varotsos, K. Alexopoulos, Thermodynamics of Point Defects and their Relation with Bulk Properties, North Holland, Amsterdam, 1986.

[25] P. Varotsos, J. Appl. Phys. 101 (2007) 123503;

P.A. Varotsos, Phys. Rev. B 75 (2007) 172107;

P. Varotsos, K. Alexopoulos, Philosophical Magazine A 42, 13 (1980).

[26] P. Varotsos, K. Alexopoulos, Phys. Rev. B 15 (1977) 4111;

P. Varotsos, W. Ludwig, K. Alexopoulos, Phys. Rev. B 18 (1978) 2683;

K. Alexopoulos, P. Varotsos, Phys. Rev. B 24 (1981) 3606.

[27] P. Varotsos, K. Alexopoulos, Physica Status Solidi (b) 102 (1980) K67.

[28] P. Varotsos, K. Alexopoulos, J. Phys. Chem. Sol. 41 (1980) 443;

P. Varotsos, K. Alexopoulos, J. Phys. Chem. Sol. 42 (1981) 409;

P. Varotsos, K. Alexopoulos, C. Varotsos, M. Lazaridou, physica status solidi (a) 88 (1985) K137.

[29] P. Varotsos, J. Phys. Chem. Sol. 42 (1981) 405;

P. Varotsos, Physica Status Solidi (b) 100 (1980) K133;

P. Varotsos, K. Alexopoulos, J. Phys. Chem. Solids 41 (1980) 1291.

[30] A. Chroneos, R.V. Vovk, Mater. Sci. Semicond. Process. 36 (2015) 179.

[31] A. Chroneos, R.V. Vovk, J. Mater. Sci. - Mater. Electron. 26 (2015) 3787.

[32] A. Chroneos, Y. Panayiotatos, R.V. Vovk, J. Mater. Sci. - Mater. Electron. 26 (2015) 2693.

[33] A. Chroneos, R.V. Vovk, J. Mater. Sci. - Mater. Electron. 26 (2015) 2113.

[34] A. Chroneos, R.V. Vovk, Solid State Ionics 274 (2015) 1.

[35] A. Chroneos, M.E. Fitzpatrick, L.H. Tsoukalas J. Mater. Sci. - Mater. Electron. 26 (2015) 3287.

[36] P. V. Varotsos, N. V. Sarlis, M. S. Lazaridou, Phys. Rev. B 59 (1999) 24;

P. Varotsos, N. Bogris, A. Kyritsis, J. Phys. Chem. Solids 53 (1992) 1007.

[37] P. A. Varotsos, Proc. Jpn. Acad., Ser. B: Phys. Biol. Sci. 82 (2006) 86;

P. A. Varotsos, N. V. Sarlis, E. S. Skordas, CHAOS 19 (2009) 023114;

P. A. Varotsos, N. V. Sarlis, E. S. Skordas, M. S. Lazaridou, J. Appl. Phys. 103 (2008) 014906

[38] P. Varotsos, K. Alexopoulos, Tectonophysics 110 (1984) 93;

P. Varotsos, K. Alexopoulos, Tectonophysics 110 (1984) 99;

E. S. Skordas, N. V. Sarlis, P. A. Varotsos, CHAOS 20 (2010) 033111;

N. V. Sarlis, E. S. Skordas, P. A. Varotsos, Phys. Rev. E 82 (2010) 021110.

[39] A.R. Denton, N.W. Ashcroft, Phys. Rev. A 43 (1991) 3161.

**Figure and Figure Caption.**

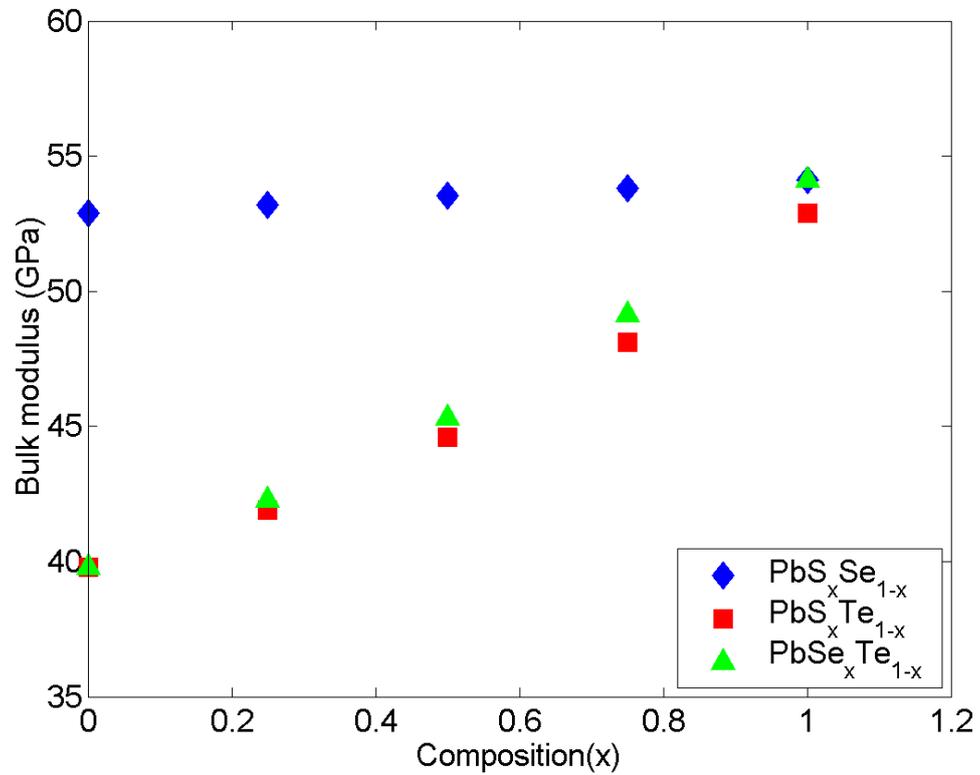

**Fig. 1**. The variation of the bulk modulus – calculated on the basis of Eq.(3)- as a function of composition " $x$ " for $PbS_xSe_{1-x}$ (blue diamonds), $PbS_xTe_{1-x}$ (red squares) and $PbSe_xTe_{1-x}$ (green triangles)

**Table 1**

Calculated lattice parameter $\alpha$ (Å) and bulk modulus $B$ (GPa) from the present study for PbS$_x$Se$_{1-x}$ compared to the experimental works and GGA calculations [19].

|  | Present study | | GGA [19] | | Experimental | |
| --- | --- | --- | --- | --- | --- | --- |
|  | $\alpha$ (Å) | $B$ (GPa) | $\alpha$ (Å) | $B$ (GPa) | $\alpha$ (Å) | $B$ (GPa) |
| PbSe | 6.1237 | 54.10 | 6.095 | 55.7 | 6.117 [33], 6.124[3,7], 6.130[15] | 54.1 [33] |
| PbS$_{0.25}$Se$_{0.75}$ | 6.077 | 53.82 | 6.05 | 56.60 |  |  |
| PbS$_{0.5}$Se$_{0.5}$ | 6.030 | 53.53 | 6.00 | 57.80 |  |  |
| PbS$_{0.75}$Se$_{0.25}$ | 5.987 | 53.20 | 5.958 | 59.74 |  |  |
| PbS | 5.936 | 52.9 | 5.895 | 60.7 | 5.929[33], 5.939[3, 7], 5.940[15] | 52.9 [33] |

**Table 2**

Calculated lattice parameter $\alpha$ (Å) and bulk modulus $B$ (GPa) from the present study for $PbS_xTe_{1-x}$ compared to the experimental works and GGA calculations [19].

| | Present study | | GGA [19] | | Experimental | |
|---|---|---|---|---|---|---|
| | $\alpha$ (Å) | $B$ (GPa) | $\alpha$ (Å) | $B$ (GPa) | $\alpha$ (Å) | $B$ (GPa) |
| PbTe | 6.4613 | 39.80 | 6.42 | 44.44 | 6.462 [33] | 39.8 [33] |
| | | | | | 6.462 [3,7], | |
| | | | | | 6.460 [15] | |
| $PbS_{0.25}Te_{0.75}$ | 6.338 | 41.90 | 6.333 | 47.51 | | |
| $PbS_{0.5}Te_{0.5}$ | 6.210 | 44.60 | 6.208 | 50.879 | | |
| $PbS_{0.75}Te_{0.25}$ | 6.076 | 48.10 | 6.067 | 55.233 | | |
| PbS | 5.936 | 52.9 | 5.895 | 60.7 | 5.929 [33] | 52.9 [33] |
| | | | | | 5.939 [3, 7] | |
| | | | | | 5.940 [15] | |

**Table 3**

Calculated lattice parameter $\alpha$ (Å) and bulk modulus $B$ (GPa) from the present study for PbSe$_x$Te$_{1-x}$ compared to the experimental works and GGA calculations [19].

|  | Present study | | GGA [19] | | Experimental | |
|---|---|---|---|---|---|---|
|  | $\alpha$ (Å) | $B$ (GPa) | $\alpha$ (Å) | $B$ (GPa) | $\alpha$ (Å) | $B$ (GPa) |
| PbTe | 6.4613 | 39.80 | 6.42 | 44.44 | 6.462 [33] 6.462[3,7], 6.460[15] | 39.8 [33] |
| PbSe$_{0.25}$Te$_{0.75}$ | 6.380 | 42.27 | 6.36 | 46.60 |  |  |
| PbSe$_{0.5}$Te$_{0.5}$ | 6.2970 | 45.31 | 6.27 | 49.26 |  |  |
| PbS$_{0.75}$Te$_{0.25}$ | 6.212 | 49.13 | 6.199 | 51.67 |  |  |
| PbSe | 6.124 | 54.10 | 6.095 | 55.74 | 6.117[33] 6.124[3, 7] 6.130[15] | 54.1 [33] |